\documentstyle[12pt]{article}
\def\gsim{\:\raisebox{-0.5ex}{$\stackrel{\textstyle>}{\sim}$}\:}
\begin{document}
\begin{titlepage}
\pagestyle{empty}
\baselineskip=21pt
\vskip .2in
\begin{center}
{\large{\bf Non-collapsing quasiparticle random phase approximation for
nuclear double-beta decay}}$^*$
 \end{center}
\vskip .1in

\begin{center}

F. Krmpoti\'{c}$^{\dagger}$, A. Mariano$^{\dagger}$

{\small\it Departamento de F\'\i sica, Facultad de Ciencias Exactas,}\\
{\small\it Universidad Nacional de La Plata, C. C. 67, 1900 La Plata, Argentina}

E.J.V. de Passos$^{\dagger\dagger}$ and A.F.R. de Toledo Piza

{\small\it Instituto de F\'\i sica, Universidade de S\~{a}o Paulo,\\
C.P. 20516, 01498 S\~{a}o Paulo, Brasil}

{\small\it and}

T.T.S. Kuo

{\small\it Physics Department, State University of New York at Stony Brook,}\\
{\small\it Stony Brook, NY 11794-3800, USA}
\end{center}
\vskip 0.5in \centerline{ {\bf Abstract} }

We show how the longstanding problem of the collapse of the
charge-exchange QRPA near the physical value of the force strength can
be circumvented. This is done by including the effect of ground state
correlations into the QRPA equations of motion. The corresponding
formalism, called renormalized QRPA, is briefly outlined and its
consequences are discussed in the framework of a schematic model for
the two-neutrino double beta decay in the $^{100}Mo \rightarrow\,
^{100}Ru$ system. The question of the conservation of the Ikeda sum rule
is also addressed within the new formalism.
\baselineskip=18pt

\vspace{0.25in}
PACS number(s): 23.40.Hc, 21.60.Jz
\bigskip

\vspace{0.5in}
\noindent
$^{*}$Work supported in part by Fundaci\'on Antorchas (Argentina).\\
$^{\dagger}$Fellow of the CONICET from Argentina.\\
$^{\dagger\dagger}$Fellow of the CNPq from Brazil.
\end{titlepage}

Double beta ($\beta\beta$) decays occur in medium-mass nuclei that are
rather far from closed shells. The nuclear structure method most
widely used in the evaluation of $\beta\beta$ rates for two-neutrino
decay mode ($\beta\beta_{2\nu}$) as well as for the neutrinoless mode
($\beta\beta_{0\nu}$) is therefore the quasiparticle random phase
approximation (QRPA) \cite{Kla94}. These calculations, in which the
$\beta\beta_{2\nu}$ matrix elements ${\cal M}_{2\nu}$ are approximated
by their $J^{\pi}=1^+$ component ({\it i.e., } ${\cal M}_{2\nu} \cong {\cal
M}_{2\nu}(J^{\pi} = 1^{+})$), explain the smallness of the measured
transition rates.
\footnote{ It was found that the contributions of the odd-parity
nuclear operators to the $\beta\beta_{2\nu}$-decay are significant
when compared with the experimental data \cite{Wil88}.}
However, the actual value of ${\cal M}_{2\nu}$ depends sensitively on
the strength $g^{pp}$ of the particle-particle force in the $S=1$,
$T=0$ channel. For realistic forces of finite range ${\cal M}_{2\nu}$
passes through zero near $g^{pp}=1$ {\it i.e., } near the physical value of
this coupling constant.  This feature makes the actual value of ${\cal
M}_{2\nu}$ rather uncertain.  What is still more distressing, is that
QRPA collapses for  $g^{pp}\gsim 1$. One may thus suspect that ${\cal
M}_{2\nu}$ goes through zero simply because the approximation breaks up.
In other words, the smallness of ${\cal M}_{2\nu}$ in
the QRPA could be just an artifact of the model. (One should remember
that in the Tamm-Dancoff approximation, {\it i.e., } in the absence of the
ground state correlations, ${\cal M}_{2\nu}$ always increases with
$g^{pp}$.) Yet, it has been pointed out more than once that the zero
of ${\cal M}_{2\nu}$ is not engendered by the collapse of the QRPA,
but arises instead from the partial restoration of the $SU(4)$ Wigner
symmetry \cite{Krm94}.

It has been shown recently that within the QRPA the above behavior of
the $2\nu$ amplitude can be summarized as \cite{Krm92}
\begin{equation}
{\cal M}_{2\nu}\cong{\cal M}_{2\nu}(g^{pp}=0)
\frac{1-g^{pp}/g^{pp}_0}{1-g^{pp}/g^{pp}_1},
~~\mbox{with}~~~g^{pp}_0\cong 1, g^{pp}_1\:\raisebox{-0.5ex}{$\stackrel{\textstyle>}{\sim}$}\:
 g^{pp}_0,
\label{1}
\end{equation}
where $g^{pp}_0$ and $g^{pp}_1$ denote respectively the zero and the
pole of ${\cal M}_{2\nu}$. Moreover, it has been suggested that within
the QRPA the $0\nu$ amplitude behaves as
\begin{eqnarray}
{\cal M}_{0\nu}&\cong& {\cal M}_{0\nu}(J^{\pi}= 1^{+};g^{pp}=0)
\frac{1-g^{pp}/g^{pp}_0}{\sqrt{1-g^{pp}/g^{pp}_1}}\nonumber\\
&+&{\cal M}_{0\nu}(J^{\pi}\neq 1^{+};g^{pp}=0)(1-g^{pp}/g^{pp}_2),
\label{2}
\end{eqnarray}
where $g^{pp}_2 \gg g^{pp}_1$ \cite{Krm92}. This means that the
$J^{\pi}=1^+$ component of ${\cal M}_{0\nu}$ exhibits the zero and the
pole at the same value of $g^{pp}$ as ${\cal M}_{2\nu}$.  
Thus, the theoretical estimation of ${\cal M}_{0\nu}$, and therefore the
determination of the limit for the effective neutrino mass $<m_{\nu}>$,
is also uncertain as that of ${\cal M}_{2\nu}$.

Several modifications of the QRPA have been proposed in order to
change the above behavior in a qualitative way, including higher order
RPA corrections \cite{Rad91}, nuclear deformation \cite{Rad93},
single-particle self-energy BCS terms \cite{Kuo92} and particle number
projection \cite{Krm93a}. Yet, none of these amendments inhibits the
collapse of the charge-exchange QRPA.  In the present work we show
that this can be achieved by including the effect of ground state
correlations in the QRPA equations of motion.  The corresponding
formalism, referred to as renormalized QRPA (RQRPA), was originally
introduced by Rowe \cite{Row68}.
It has been used recently by Catara {\it et al., } \cite{Kar93,Cat94} in the
evaluation of the charge transition densities and properties of the
charge-conserving collective states. We briefly outline below the
RQRPA formalism for charge-exchange excitations, and discuss it within
a schematic model for ${\cal M}_{2\nu}$.

We begin by defining excited states $|\lambda J\rangle$ that are built by
the action of the charge-exchange operators
\begin{equation}
\Omega^{\dag}(\lambda J) = \sum_{pn} \left[ X_{pn}(\lambda J)
 A^{\dag}_{pn}(J) - Y_{pn}(\lambda J) A_{pn}(\bar J)\right],
\label{3}\end{equation}
on the correlated ground state $|0\rangle$. Here $A^{\dag}_{pn}(J)
=[\alpha^{\dagger}_p\alpha^{\dagger}_n]^{J}$, and $\alpha^{\dagger}_p$ and
$\alpha^{\dagger}_n$ are quasiparticle creation operators for protons and
neutrons.  The amplitudes $X$ and $Y$, the eigenvalues $\omega_\lambda$ and
$|0\rangle$ are obtained from the equations of motion (EM)
\begin{equation}
\langle 0|\left[\delta\Omega(\lambda \bar{J}),H,\Omega^\dagger (\lambda J)
\right]^0|0 \rangle =
\omega_\lambda \langle 0|\left[\delta\Omega(\lambda \bar{J}),
\Omega^\dagger (\lambda J) \right]^0 |0\rangle,
\label{4} \end{equation}
with the condition
\begin{equation}
\Omega(\lambda J)|0\rangle = 0 ,\mbox{for all}~ \lambda, J.
\label{5} \end{equation}
The usual QRPA equations result from (\ref{4}) when $|0\rangle$ is
approximated by the $BCS$ ground state $|BCS\rangle$ and (\ref{5}) is
ignored.  In the RQRPA one takes the ground state correlations (GSC)
introduced by (\ref{5}) in the EM (\ref{4}) partially into account.  First
note that we have now
\begin{equation}
\hat{J}^{-1} \langle 0|\left[A_{pn}(\bar{J}),A^{\dag}_{p'n'}(J')\right]^0
|0\rangle  = \delta_{pp'}\delta_{nn'} \delta_{JJ'} D_{pn},
\label{6}\end{equation}
with
\begin{equation}
D_{pn}= \hat{J}^{-1}\langle 0|\left[A_{pn}(\bar{J}),A^{\dag}_{pn}(J)\right]^0
|0\rangle  = 1 - {\cal N}_p -{\cal N}_n,
\label{7} \end{equation}

where $\hat{J}\equiv\sqrt{2J+1}$ and  ${\cal N}_p$ (${\cal N}_n$) are the
proton (neutron) quasiparticle occupations
\begin{equation}
{\cal N}_t= \hat{j}_t^{-1} \langle 0|[\alpha^\dagger_t\alpha_{\bar{t}}]^0
|0\rangle.
\label{8} \end{equation}
The label $t$ stands for $p$ and $n$.

We define next ``renormalized'' two quasiparticle operators as
\begin{equation}
{\cal A}^{\dag}_{pn}(J)=A^{\dag}_{pn}(J)D_{pn}^{-1/2},
\label{9} \end{equation}
which satisfy the relation
\begin{equation}
\hat{J}^{-1}\langle 0|\left[{\cal A}_{pn}(\bar{J}),
{\cal A}^{\dag}_{p'n'}(J')\right]^0|0\rangle  = 
\delta_{pp'}\delta_{nn'} \delta_{JJ'}.
\label{10} \end{equation}
The crucial RQRPA assumption is the generalized quasiboson approximation
\begin{equation}
\hat{J}^{-1}\left[{\cal A}_{pn}(\bar{J}),{\cal A}^\dagger_{p'n'}(J')\right]^0
\cong \hat{J}^{-1}\langle 0|\left[{\cal A}_{pn}(\bar{J}),
{\cal A}^\dagger_{p'n'}(J')\right]^0|0\rangle
= \delta_{pp'}\delta_{nn'} \delta_{JJ'},
\label{11} \end{equation}
The RQRPA equations follow straightforwardly after replacing
$A^{\dag}_{pn}(J)$ by ${\cal A}^{\dag}_{pn}(J)$ in the expression
for $\Omega^\dagger(J)$ and using (\ref{11}) in the EM (\ref{4}).
We get in this way \cite{Row68,Cat94}
\begin{eqnarray}
\left(\begin{array}{ll} {\sf A}(J) & {\sf B}(J)
\\ {\sf B}^{*}(J) & {\sf A}^{*}(J)\end{array}\right)
\left(\begin{array}{l} {\sf X}(\lambda J) \\ {\sf Y}(\lambda J)
\end{array}\right)
= \omega_{\lambda J}
\left(\begin{array}{l} ~{\sf X}(\lambda J) \\ - {\sf Y}(\lambda J)
\end{array}\right),
\label{12} \end{eqnarray}
where
\begin{equation}
{\sf X}_{pn}(\lambda J)\equiv X_{pn}(\lambda J)  D_{pn}^{1/2}~~~
\mbox{and}~~~{\sf Y}_{pn}(\lambda J)\equiv Y_{pn}(\lambda J)  D_{pn}^{1/2},
\label{13} \end{equation}
are the renormalized amplitudes. The submatrices ${\sf A}(J)$ and
${\sf B}(J)$ are found as
\begin{eqnarray}
{\sf A}_{pn,p'n'}(J) & = & (\epsilon_p+\epsilon_n)\delta_{pp'}\delta_{nn'}
 + D_{pn}^{1/2}\left[F(pn,p'n',J) (u_pv_nu_{p'}v_{n'}+v_pu_nv_{p'}u_{n'} )
\right.\nonumber \\
&+ &\left. G(pn,p'n',J) (u_pu_nu_{p'}u_{n'}+v_pv_nv_{p'}v_{n'})\right]
D_{p'n'}^{1/2}, \nonumber \\
{\sf B}_{pn,p'n'}(J)& = & D_{pn}^{1/2}\left[F(pn,p'n',J)
(v_pu_nu_{p'}v_{n'}+u_pv_nv_{p'}u_{n'}) \right. \nonumber \\
& - &\left. G(pn,p'n',J)(u_pu_nv_{p'}v_{n'} + v_pv_nu_{p'}u_{n'})\right]
D_{p'n'}^{1/2},
\label{14} \end{eqnarray}
where $F$ and $G$ are the usual particle-hole (PH) and particle-particle
(PP) coupled two-particle matrix elements.

The QRPA equations are recovered from (\ref{13}) and (\ref{14}) by taking
$D_{pn}=1$.  Within the RQRPA one first solves (\ref{5}) in the
quasiboson approximation \cite{Row68}. The RQRPA ground state then reads
\begin{equation}
|0\rangle = N_0 e^{{\cal S}} |BCS\rangle,
\label{15} \end{equation}
with
\begin{equation}
{\cal S} = \frac{1}{2} \sum_{pnp'n'J} \hat{J}^{-1}\left[{\sf C}_{pnp'n'}(J)
{\cal A}^{\dag}_{pn}(J){\cal A}^{\dag}_{p'n'}(J)\right]^0 .
\label{16} \end{equation}
From (\ref{5}) it turns out that the matrix ${\sf C}$ is the
solution of 
\begin{equation}
\sum_{pn} {\sf X}_{pn}^*(\lambda J){\sf C}_{pnp'n'}(J)
= {\sf Y}_{p'n'}^*(\lambda J), ~\mbox{for all $\lambda$, $J$}.
\label{17} \end{equation}
Finally, by making use of this equation one finds the quasiparticle
occupations
\begin{equation}
{\cal N}_p=\sum_{\lambda J n'}\hat{J}^2\hat{j}_p^{-2}|
{\sf Y}_{pn'}(\lambda J)|^2;~~
{\cal N}_n=\sum_{\lambda J p'}\hat{J}^2\hat{j}_n^{-2} |
{\sf Y}_{p'n}(\lambda J)|^2.
\label{18} \end{equation}
The value of $D_{pn}$ follows from (\ref{7}) and (\ref{18}).

To evaluate the transition matrix elements for the $\beta^{\mp}$ decays
\begin{equation}
\langle\lambda J||{\cal O}(J;\pm) ||0\rangle =
\langle 0|\left[\Omega(\lambda \bar{J}),{\cal O}(J;\pm)\right]^0|0\rangle,
\label{19} \end{equation}
with
\begin{equation}
{\cal O}(J;\pm)=\sum_iO(J;i)t_{\pm}(i),
\label{20} \end{equation}
we only need their two quasiparticle components
\begin{equation}
{\cal O}(J;\pm) \:\raisebox{-0.5ex}{$\stackrel{\textstyle.}{=}$}\:
\sum_{pn}\left[\Lambda^0_{pn}(J;\pm) A^{\dag}_{pn}(J)
+ (-)^J\Lambda^{0*}_{pn}(J;\mp)A_{pn}(\bar J)\right],
\label{21} \end{equation}
where
\begin{eqnarray} 
\Lambda^0_{pn}(J;+) & = &- \hat{J}^{-1}u_pv_n \langle p||O(J)||n\rangle,
\nonumber \\
~\Lambda^0_{pn}(J;-) & = &(-)^J\hat{J}^{-1}u_nv_p \langle p||O(J)||n\rangle^*.
\label{22} \end{eqnarray}
From (\ref{3}) and (\ref{22}) one gets
\[
\langle\lambda J||{\cal O}(J;\pm)||0\rangle =\hat{J}
\sum_{pn}\left[\Lambda^0_{pn}(J;\pm) {\sf X}^*_{pn}(\lambda J)
+(-)^J \Lambda^{0*}_{pn}(J;\mp){\sf Y}^*_{pn}(\lambda J)\right]D_{pn}^{1/2}.
\]
The corresponding total strengths are
\begin{equation}
S(J;\pm)=\hat{J}^{-2}
\sum_{\lambda}|\langle\lambda J||{\cal O}(J;\pm)||0\rangle|^2.
\label{23} \end{equation}

Within the RQRPA the BCS equations have to be solved subject to the
condition that $|0\rangle$ has on the average the correct number of
particles. This requirement gives
\begin{equation}
{\sf N}_t=\sum_t\hat{j}_t^2[v_t^2+(1-2v_t^2){\cal N}_t],
\label{24} \end{equation}
${\sf N}_p$ and ${\sf N}_n$ being the number of active protons and neutrons
in solving the gap equations.

We conclude the presentation of the formalism by noting that: (a) when
the factors $D_{pn}$, which are functions of the amplitudes ${\sf Y}$
are substituted into the renormalized matrices ${\sf A}$ and ${\sf
B}$, (\ref{12}) becomes a nonlinear system of coupled equations for
the ${\sf X}$ and ${\sf Y}$ amplitudes; and (b) these equations have
to be solved self-consistently together with the new $BCS$ conditions
(\ref{24}).  This is the price to be paid in order to take into account
the GSC within the QRPA problem in an appropriate way.

We will resort now to the simplest version of the QRPA for the
$\beta\beta$-decay, called the single mode model (SMM), in which
a single RPA equation is solved with two BCS vacua
\cite{Hir90}, and only one intermediate state $J^{\pi}=1^+$ enters into
the play \cite{Krm92}. Eqs. (\ref{14}) read in this case
\begin{eqnarray*}
A_{pn} & \equiv & \omega_0+\rho_p\rho_n\left[(u_p^2v_n^2
+ \bar{v}_p^2 \bar{u}_n^2)F(pn;1)
+ (u_p^2 \bar{u}_n^2 + \bar{v}_p^2 v_n^2) G(pn;1)\right]D_{pn},\nonumber \\
B_{pn} & \equiv & 2\rho_p\rho_n \bar{v}_p \bar{u}_n v_n u_p
[ F(pn;1)- G(pn;1)]D_{pn},
\end{eqnarray*}
where $\omega_0=-[G(pp;0)+G(nn;0)]/4$ is the unperturbed energy.  The
unbarred (barred) quantities indicate that the quasiparticles are
defined with respect to the initial (final) nucleus;
$\rho_p^{-1}=u_p^{2}+\bar{v}_p^{2}$, $\rho_n^{-1}=\bar{u}_n^{2}+v_n^{2}$.  All the
remaining notation is self explanatory. The perturbed energy and
$D_{pn}$ are obtained by solving self-consistently the set of
equations:
\[
\omega =\sqrt{A_{pn}^2-B_{pn}^2},~~~D_{pn}=1-f\frac{A_{pn}-\omega}{2\omega},
~~~v_t^2=\frac{{\sf N}_t f - 3(1 - D_{pn})}{f\hat{j}_t^2- 6(1 - D_{pn})},
\]
with $f\equiv 3(\hat{j}_p^{-2}+\hat{j}_n^{-2})$.

The transition $\beta\beta_{2\nu}$ matrix element is
\begin{equation}
{\cal M}_{2\nu}= {\cal M}_{2\nu}^0 D_{pn}
\left(\frac{\omega_0} {\omega}\right)^{2}
\left(1+\frac{G(pn;1)D_{pn}}{\omega_0}\right),~~~
{\cal M}_{2\nu}^0=\frac{\rho_p\rho_n\bar{v}_p\bar{u}_nv_nu_p}
{\omega_0}|\langle p||\sigma||n\rangle|^2
\label{25}\end{equation}
with ${\cal M}_{2\nu}^0$ being the corresponding unperturbed (BCS) value.

Numerical calculations have been performed for the $^{100}Mo
\rightarrow\,^{100}Ru$ system, where the appropriate intermediate
state is $[0g_{7/2}(n)0g_{9/2}(p)]^1$, and ${\sf N}_p=2$ and ${\sf
N}_n=2$ (${\sf N}_p=4$ and ${\sf N}_n=0$) for the initial (final)
state.  We have used a $\delta$-force (in units of $MeV fm^{3}$):
$V=-4\pi({\it v}_{s}P_{s}+{\it v}_{t}P_{t})\delta(r)$, with different
strength constants ${\it v}_{s}$ and ${\it v}_{t}$ for the PH, PP and
pairing channels. Thus, instead of the parameter $g^{pp}$ we use here
the ratio ${\rm t}={\it v}_t^{pp}/{\it v}_s^{pair}$, whose physical
value is ${\rm t}\cong 1.5$. The remaining parameters for the SMM have
been taken to be ${\it v}_s^{ph}=55$, ${\it v}_t^{ph}=92$ and ${\it
v}_s^{pair}=55$ \cite{Krm94}. The results obtained within the QRPA
(dashed lines) and the RQRPA (solid lines) for $\omega$ and for ${\cal
M}_{2\nu}$ are shown in Fig. 1. As expected, the QRPA collapses
close to ${\rm t}= 1.5$.  Contrarily, in the RQRPA the energy decreases
asymptotically when ${\rm t}\rightarrow \infty$.
For the sake of comparison, in the same figure, are also presented the results
for the energy of the lowest $J^{\pi}=1^+$ state and for the $2\nu$ matrix
element of a full QRPA calculation (dotted lines), as described in ref.
\cite{Krm94}.
This calculation, that involves an eleven dimensional model space, both for
protons and neutrons, also collapses.
(It is very gratifying that the simple formula (\ref{25}) contains the main
physics involved in such a relative sizable calculations.)

In summary, we have investigated the importance of GSC effects on the solutions
of the EM for charge-exchange excitations in the renormalized QRPA. The SMM
shows that, contrarily to what happens in the usual QRPA, the inclusion of
the GSC in the EM avoids collapse for physical values of the PP coupling
strength. However, the amplitude ${\cal M}_{2\nu}$ still passes through zero
in the RQRPA, although at somewhat higher value of ${\rm t}$ (or $g^{pp}$).
It is also evident that, in the QRPA, the physical mechanisms  responsible
for the zero and the collapse of  ${\cal M}_{2\nu}$ are not the same.
The behavior of this amplitude in the RQRPA is not anymore delineated by
Eq. (\ref{1}), and the dependence of the calculated $\beta\beta_{2\nu}$
transition rates on $g^{pp}$ is weakened.
In view of Eqs. (\ref{1}) and (\ref{2}), all that was just said
for the $2\nu$ mode can be extrapolated also to the $0\nu$ mode.  It
is well known that the contributions of intermediate states with
$J^{\pi}\ne 1$ are quite sizeable in the neutrinoless decay for
physical value of $g^{pp}\cong 1$, where it is very likely that the
${\cal M}_{0\nu}(J^{\pi}=1^+)$ goes to zero even in the RQRPA case.
But as the dynamical calculation does not collapse any more, we could now have
more confidence in establishing the upper limit for the neutrino mass.
Thus, the effect of the GSC in the EM appear in this context as
particularly relevant.
We have also found that a full RQRPA calculation for the ${\cal M}_{2\nu}$
amplitude agrees qualitatively with the SMM estimate.
But, in analyzing the Ikeda sum rule $S(J^{\pi}=1^+;+)-S(J^{\pi}=1^+;-)=N-Z$,
we discovered that it is not fulfilled  within the RQRPA.
In fact, the deviations from this condition
grow as the GSC increase (or as the PP strength parameter increases).
On the other hand, we have verified numerically that the similar requisite for
the Fermi transitions is fulfilled in our formalism, when only the states
$J^{\pi}=0^+$ are considered in the Eqs. (\ref{18}). It should be
stressed that the constraints (\ref{24}) plays a crucial role regarding this
point.
When the usual BCS constraint on the number of particles is used \cite{Kar93},
the sum rule for the Fermi transitions is never fulfilled.
That the Ikeda sum rule is necessarily violated in the RQRPA, when the usual
BCS occupation numbers are employed, is seen immediately from the relation
\[
S(J^{\pi}=1^+;+)-S(J^{\pi}=1^+;-)
=\frac{1}{3}\sum_{pn}|<p||\sigma||n>|^2(v_n^2-v_p^2)D_{pn},
\]
which yields $N-Z$ only when $D_{pn}\equiv 1$.
Why the Ikeda sum rule is not satisfied, even when the condition (\ref{24}) is
adopted, is still an open question.
In summary, we feel that, before a quantitative comparison of the calculations
with the experimental data could be done, the behavior of the sum rules in the
RQRPA should be thoroughly elucidated and this is our next goal.

Finally, it should be mentioned that, after our work has been completed, we
have learned that a similar study has been performed by Toivanen and
Suhonen \cite{Toi95}.

\newpage

\bigskip

\begin{figure}
\begin{center} { \bf Figure Captions} \end{center}
\caption{ \protect \footnotesize
Energies $\omega$ (in $MeV$) of the lowest $1^+$ state of $^{100}Tc$
and the matrix elements ${\cal M}_{2\nu}$ (in $[MeV]^{-1}$) for the
$^{100}Mo \rightarrow\, ^{100}Ru$ system. The single mode model
results are indicated by the dashed lines for the QRPA and by the solid
lines for the RQRPA.
The results of a full QRPA calculation \protect \cite{Krm94}
are represented by dotted lines.
\label{fig1}}
\end{figure}
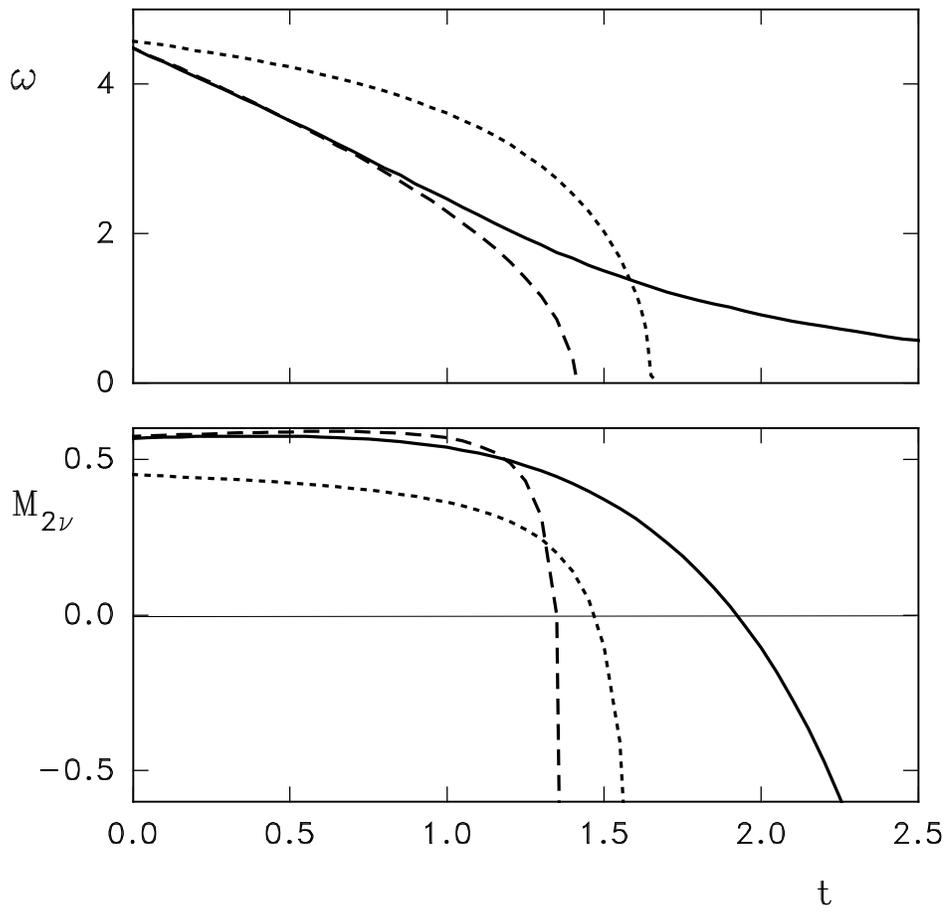
\end{document}